\documentclass{ws-procs9x6}
\usepackage{epsfig}
\usepackage{amssymb,amsbsy,amsmath,amsfonts}

\def\beq{\begin{equation}}
\def\eeq{\end{equation}}
\def\bea{\begin{eqnarray}}
\def\eea{\end{eqnarray}}
\def\beqa{\begin{equation}\begin{array}{l}}
\def\eeqa{\end{array}\end{equation}}
\def\eqlab#1{\label{eq:#1}}
\def\figlab#1{\label{fig:#1}}
\def\tablab#1{\label{tab:#1}}

\def\barr{\left(\begin{array}{c}}
\def\earr{\end{array}\right)}
\def\bmat{\left(\begin{array}{cc}}
\def\emat{\end{array}\right)}
\def\Eqref#1{Eq.~(\ref{eq:#1})}

\def\Figref#1{Fig.~\ref{fig:#1}}

\def\Tabref#1{Table \ref{tab:#1}}

\def\sla#1{#1  \!\!\!\!\slash}

 \def\Ga{{\it\Gamma}}
\def\de{\delta} \def\De{\Delta}
\def\vDe{\varDelta}
  \def\eps{\epsilon}

 \def\La{{\Lambda}}

 \def\Si{{\it\Sigma}}
  
\def\w{\omega}

\def\pa{\partial}

\def\pa{\partial}

\def\nn{\nonumber}

\def\lag{{\mathcal L}}

\def\mathscr{\mathcal}

\def\3d{3-D}

\def\ol#1{\overline{#1}}

\begin{document}

\title{Chiral Effective Field Theory in the $\Delta$-resonance region}

\author{Vladimir Pascalutsa\footnote{Present address: {\it ECT*, Villa Tambosi, Strada delle Tabarelle 286,
I-38050 Villazzano-Trento, Italy}. E-mail: {\it vlad@ect.it}\ .}
}

\address{Physics Department, The College of William \& Mary,
Williamsburg, VA 23187, USA\\
Theory Center, Jefferson Lab, 12000 Jefferson Ave, Newport News,
VA 23606, USA}

\begin{abstract}
I discuss the problem of constructing an effective low-energy theory
in the vicinity of a resonance or a bound state. The focus is on the
example of the $\Delta(1232)$, the lightest resonance in the nucleon
sector. Recent developments of the chiral effective-field
theory in the $\Delta$-resonance region are briefly reviewed.
I conclude with a comment on the merits of the manifestly covariant
formulation of chiral EFT in the baryon sector.
\end{abstract}


\bodymatter

\section{Introduction}\label{sec:sec1}

Chiral Perturbation Theory ($\chi$PT) 
provides a systematic field-theoretic framework
for the description of the low-energy strong interaction. The fundamental
degrees of freedom in $\chi$PT are therefore the low-energy hadron excitations,
such as the Goldstone bosons (GBs) of chiral symmetry breaking, nucleons 
and a few others. 
The corresponding hadron fields appear in the effective Lagrangian 
which can be organized
in powers of derivatives of the GB fields, or schematically as: 
\beq
{\cal L}(\pi, N, \ldots) = \sum_{n} {\mathcal L}^{(n)}=
\sum_{n} {\cal O}_n(c_i) 
\frac{(\pa \pi)^n}{\La^n}
\eeq
where
$O_n$ are some field operators which may contain GB fields
but not their derivatives. The all-possible field operators,
constrained by chiral and other
symmetries, appear with the free parameters, $c_i$,
the so-called low energy constants (LECs).
The mass scale $\La$ is the heavy scale
which sets the upper limit of applicability of $\chi$PT
and is believed to be of order of $1$ GeV ---  the scale
of spontaneous chiral symmetry breaking that led to
the appearance of the GBs.

 This expansion
of the Lagrangian translates into a low-energy expansion
of the $S$-matrix, schematically:
\beq
S = \sum_{n}\, A_n( c_i) \, \frac{ p^n}{\La^n}
\eeq
where $A$'s are amplitudes which depend on LECs, and $p$ denotes the
momentum of the GBs.  
The anzatz\cite{Weinberg:1978kz} 
is that the same expansion can (one day) be obtained
directly from QCD, provided the LECs are matched onto the
QCD parameter: $c_n = c_n (\La_{QCD})$. In the absence
of the correspondent calculation in QCD, the best one can do is
to match (or, fit) the LECs to experimental data, making sure
that they take reasonable (or, natural) values such that
the above expansion is convergent.  

One case where the convergence of the $\chi$PT expansion is
immediately questioned is the case of hadronic bound states and
resonances. In the presence of a bound state or a resonance
the low-energy expansion of the $S$-matrix
goes as  
\beq
S \sim \sum_{n} A_n \, \left(
\frac{p}{ { \Delta E} }\right)^n,
\eeq
where ${\Delta E}$ is the excitation (binding) energy of 
the resonance (bound state). Thus, the limit
of applicability of $\chi$PT is limited not by $\La\sim 1$ GeV but by
the characteristic energy scale $\De E$ of the closest bound 
or excited state.
Furthermore, in the vicinity of a bound state or a resonance
the $S$-matrix has a pole, which cannot be reproduced in a
purely perturbative expansion in energy that is
utilized in $\chi$PT.

This problem arises in various contexts, ranging from
pion-pion scattering\cite{Caprini:2005zr} 
to halo nuclei\cite{vanKolck:2004te}. Some 
are being discussed
at this meeting (e.g.,  resonances in the $\pi\pi$
system\cite{Leut}, or bound states and resonances
 in the few-nucleon system\cite{Hammer:2006qj}).
In this contribution I focus on the $\pi N$ system where the first resonance
is the $\De(1232)$.

The $\De$ resonance is an ideal study case for the problem of resonances in
$\chi$PT.
It is relatively light, with the excitation energy of $\vDe\equiv M_\De -M_N
\approx 300$ MeV, elastic, and well separated from the other 
nucleon resonances. It is also a very prominent resonance 
and plays an important
role in many processes, including astrophysical ones. It is, for instance,
responsible for the damping of the high-energy cosmic rays by the
cosmic microwave background, the so-called GZK cutoff.
Let us therefore look at 
the description of this resonance in the framework of
chiral effective-field theory ($\chi$EFT). 

\section{Power counting(s) for the $\De$ resonance}

Imagine the Compton scattering on the nucleon.
The total cross-section of this process, as the function
of photon energy $\w$, is shown in 
\Figref{totalcs}. In this case we are 
able to examine the entire energy range,
starting with $\w=0$, through the pion production
threshold $\w\simeq m_\pi$ and into the resonance region
$\w\sim \vDe$. 
\begin{figure}[t,b]
\centerline{  \epsfxsize=8cm
  \epsffile{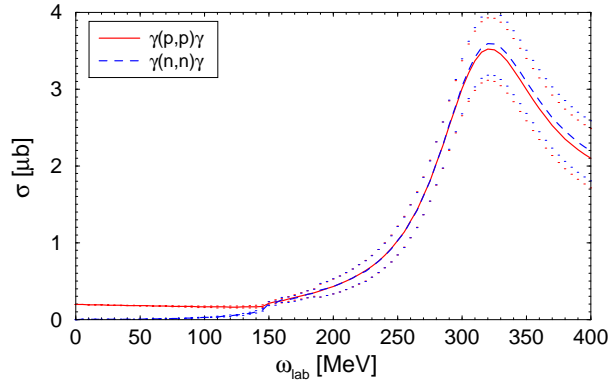} 
}
\caption{(Color online) 
Total cross-section of the Compton scattering on the
nucleon (proton -- red solid curve, neutron -- blue dashed curve),
as the function of the incident photon lab energy. The curves
are obtained in a $\chi$EFT calculation\cite{Pascalutsa:2002pi}.}
\figlab{totalcs}
\end{figure}

At energies up to around the pion production
threshold the cross section shows a smooth behavior
which can reproduced by a low energy expansion.
In this region the $\De$-resonance
 can be ``integrated out'', as its tail contribution can be 
mimicked by the terms already present in the $\chi$PT Lagrangian
with nucleons only\cite{GSS89}.  

Higher in energy, however, the rapid energy variation induced
by the resonance pole is not reproducible by a naive low-energy
expansion. Obviously, to describe this behavior it is necessary to
introduce the $\De$ as an explicit degree of freedom\cite{JeM91a}, hence
include a corresponding field in the effective chiral Lagrangian.
The details of how this is done have recently been reviewed 
in\cite{Pascalutsa:2006up}. 

Once the $\De$ appears in the Lagrangian the question is how
to power-count its contributions. In $\chi$EFT with
pions and nucleons alone the power-counting index of a graph
with $L$ loops, $N_\pi$ ($N_N$) internal pion (nucleon)
lines, and $V_k$ vertices from $k$th-order Lagrangian
is found as
\beq
\eqlab{chptindex}
n_{\chi \mathrm{PT}} =  4 L  - 2 N_\pi - N_N + \sum_k k V_k\,.
\eeq
What about the graphs with the $\De$, such as those depicted
in \Figref{ODR} ? Their power counting turns out to be 
dependent on how one weighs the excitation energy $\vDe$
in comparison with the other mass scales of the theory. 
In this case we have the soft momentum $p$ (or, $\w$),
the pion mass $m_\pi$, and heavy scales which we collectively
denote $\La$. 

The Small Scale Expansion\cite{HHK97} (SSE) 
counts all light scales equally: $p\sim m_\pi \sim \vDe$.
The small parameter is then:
\beq
\eps = \left\{\frac{p}{\La},\, \frac{m_\pi}{\La},\, 
\frac{\vDe}{\La}\right\}\,.
\eeq
An unsatisfactory feature of such a democratic counting
($\eps$-expansion) is that the $\De$-resonance 
contributions are always estimated to be of the
same size as the nucleon contributions. 
As we have seen from \Figref{totalcs}, in reality
the resonance contributions 
are suppressed at low energies while being  dominant
in the resonance region. 
Therefore the power counting
{\it overestimates} the $\De$-contributions
at lower energies and {\it underestimates} them
at the resonance energies. 
Despite this flaw, the SSE has been widely and quite successfully 
used to account the $\De$ contributions
below the resonance\cite{Hemmert:2003cb,Bernard:2005fy,HWGS05,Gellas:1998wx,Gail,Hildebrandt:2003fm}.
No applications of this scheme
to calculation of observables in the resonance region have
been reported yet.
\begin{figure}[t,b]
\centerline{  \epsfxsize=8cm
  \epsffile{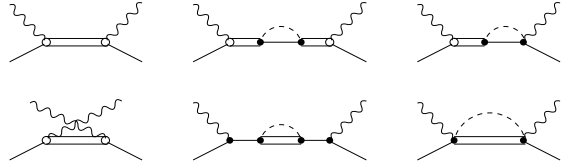} 
}
\caption{Examples of the one-Delta-reducible (1st row) and
 the one-Delta-irreducible (2nd row) graphs in Compton scattering. }
\figlab{ODR}
\end{figure}

A more adequate power counting is achieved by separating out
the resonance energy, e.g., by maintaining the following
scale hierarchy $m_\pi\ll \vDe\ll \La$
in the power-counting scheme\cite{Pascalutsa:2002pi,Hanhart:2002bu}. 
In the so-called 
{\it ``$\de$ expansion''}\cite{Pascalutsa:2002pi}
this is done by introducing a small parameter $\de=\vDe/\La$,
and then counting $m_\pi/\La$ as $\de^2$. The power 2 is chosen
here because it is the closest integer representing the ratio
of these scales in the real world. 

\begin{table}[h]
{\centering
\begin{tabular}{|c||c|c|}
\hline
EFT &  $\quad p\sim m_\pi \quad$ & $\quad p\sim \Delta \quad$\\
\hline
$\sla{\De}$-$\chi$PT & ${\mathcal O}(p)$        & ${\mathcal O}(1)$\\
 $\eps$-expansion  & ${\mathcal O}(\epsilon) $ & ${\mathcal O}(\epsilon) $\\
$\delta$-expansion & ${\mathcal O}(\delta^2) $ & ${\mathcal O}(\delta) $\\
\hline
\end{tabular}  \par }
\caption{The counting of momenta in 
the three different $\chi$EFT expansions.}
\tablab{compare}
\end{table}
Obviously, the power 
counting of the $\De$ contributions then becomes dependent on the
energy domain: in the {\it low-energy region} ($p\sim m_\pi$)
and the {\it resonance region} ($p\sim \vDe$), the momentum
counts differently, see \Tabref{compare}.
This dependence 
most significantly affects the counting of the 
one-Delta-reducible (ODR) graphs. The 1st row of graphs
in \Figref{ODR}
illustrates examples of the ODR graphs for the Compton scattering
case. These graphs are all characterized by having
a number of ODR propagators, each going as
\beq
S_{ODR}\sim \frac{1}{s-M_\De^2} \sim \frac{1}{2M_\De}\frac{1}{p-\vDe}\, ,
\eeq
where $s=M_N^2 + 2M_N\w$ is the Mandelstam variable, and
the soft momentum $p$ in this case given by the photon energy.
In contrast, the nucleon propagator in analogous graphs would go
simply as $S_N\sim 1/p$.
Therefore, in the low-energy region, the $\De$ and nucleon 
propagators would
count respectively as ${\mathcal O}(1/\de)$ and ${\mathcal O}(1/\de^2)$, 
the $\De$ being suppressed by one power of the small parameter
as compared to the nucleon. 
In the resonance region, the ODR graphs obviously 
all become large. Fortunately they all can be subsumed, leading
to ``dressed'' ODR graphs with a definite power-counting index.
Namely, it is not difficult to see that the resummation of
the classes of ODR graphs results 
in ODR graphs with only a single ODR propagator of
the form
\beq
S_{ODR}^\ast = \frac{1}{S_{ODR}^{-1} - \Sigma }
\sim \frac{1}{p-\vDe-\Sigma}\,,
\eeq
where $\Sigma$ is the $\De$ self-energy.
The expansion of the self-energy begins with $p^3$, and hence
in the low-energy region 
does not affect the counting of the $\De$ contributions. However,
in the resonance region the self-energy not only ameliorates
the divergence of the ODR propagator at $s=M_\De^2$ but also
determines power-counting index of the propagator.
Defining the $\De$-resonance region formally as the region of $p$
where
\beq
|p-\vDe | \leq \de^3 \La\,,
\eeq
we deduce that an ODR propagator, in this region, counts
as ${\mathcal O}(1/\de^3)$. Note that the nucleon propagator in
this region counts as ${\mathcal O}(1/\de)$, hence is
suppressed by two powers as compared to ODR propagators.
Thus, within the power-counting scheme we have the mechanism for
estimating correctly  the relative size
of the nucleon and $\De$ contributions in the two energy domains.
In \Tabref{counting} we summarize the counting of the nucleon,
ODR, and one-Delta-irreducible (ODI) propagators in 
both the $\eps$- and $\de$-expansion.
\begin{table}[htb]
{\centering \begin{tabular}{||c|c||c|c||}
\hline
&  $\eps$-expansion & 
\multicolumn{2}{|c||}{$\de$-expansion} \\
\cline{2-4} 
  & $p/\La_{\chi SB}\sim \eps$  &  $p\sim m_\pi$ & $p\sim \vDe$ \\
\hline
$\,S_N\,$ &  $1/\eps$ & $1/\de^2$ & $1/\de$\\
$\,S_{ODR}\,$ & $1/\eps$ & $1/\de$ & $1/\de^3$\\
$\,S_{ODI}\,$ & $1/\eps$ & $1/\de$ & $1/\de $ \\
\hline
\end{tabular} \par }
\caption{The counting for the nucleon, one-Delta-reducible (ODR), and
one-Delta-irreducible (ODI) propagators in the two different expansion
schemes. The counting in the $\de$-expansion depends on the energy domain.}
\tablab{counting}
\end{table}
\newline
\indent
We conclude this discussion by giving the general
formula for the power-counting index in the $\de$-expansion.
The power-counting index, $n$, of a given graph
simply tells us that the graph is of the size of ${\mathcal O}(\de^n)$.
For a graph with $L$ loops, $V_k$ vertices of
dimension $k$, $N_\pi$ pion propagators, $N_N$
nucleon propagators, $N_\De$ Delta propagators, $N_{ODR}$
ODR propagators and  $N_{ODI}$ ODI propagators 
(such that $N_\De=N_{ODR}+N_{ODI}$) the index is
\beq
\eqlab{PCindex}
n = \left\{ \begin{array}{cc} 2 n_{\chi \mathrm{PT}} - N_\De\,, 
& p\sim m_\pi ; \nn\\
n_{\chi \mathrm{PT}} - 3N_{ODR} - N_{ODI}\,, & p\sim \De, \end{array}\right.
\eeq
where $ n_{\chi \mathrm{PT}}$, given by \Eqref{chptindex}, 
is the index of the graph in $\chi$PT with no $\De$'s.

In the following I show a few applications of the $\de$ expansion
to the calculation of processes in the $\De$ resonance region.

\section{Pion-nucleon scattering}

The pion-nucleon ($\pi N$) scattering amplitude at leading order
in the $\de$-expansion in the resonance region, is given 
by the graph $(LO)$ in \Figref{pin2NLO}. This is an example
of an ODR graph and thus the $\De$-propagator counts
as $\de^{-3}$. The leading-order vertices are from $\lag^{(1)}$
and since $p\sim\de$, the whole graph is ${\mathcal O}(\de^{-1})$. 
\begin{figure}
\centerline{  \epsfxsize=7.5cm
  \epsffile{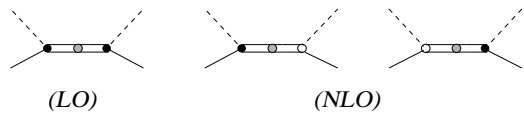} 
}
\caption{The leading and next-to-leading order graphs of the 
$\pi N$-scattering amplitude.}
\figlab{pin2NLO}
\end{figure}
\newline
\indent
At the NLO, the graphs labeled $(NLO)$ \Figref{pin2NLO} begin 
to contribute. The $\pi N\De$ vertices denoted by dots 
stand for the $h_A$ coupling from $\lag^{(1)}$ and the circles
for the $h_1$ $\pi N\De$ coupling\cite{Pascalutsa:2006up} from $\lag^{(2)}$.
The NLO graphs are thus ${\mathcal O}(\de^0)$. 
The graphs containing the loop correction
to the vertex, as well as the nucleon-exchange graphs, 
begin to contribute at N$^2$LO $[{\mathcal O}(\de)]$.

The ODR graphs at NLO contribute only to the $P_{33}$
partial wave and this contribution can
conveniently be written in terms of the following partial-wave
`$K$-matrix':
\beq
\eqlab{kmat33}
K_{P33}=-\frac{1}{2}\frac{\Gamma(W)}{W-M_\De} \,,
\eeq 
where $W=\sqrt{s}$ is the total energy and $\Gamma$ is an
energy-dependent width, which arises from the $\De$ self-energy.
At this stage it is already taken into account that the
real part of the self-energy will lead to the mass and field renormalization
and otherwise are of N$^2$LO. Thus, only the imaginary part of the
self-energy affects the NLO calculation.
\begin{figure}
\centerline{  \epsfxsize=11cm
  \epsffile{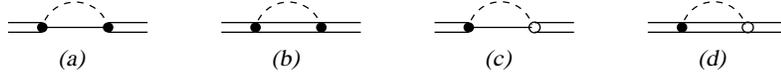} 
}
\caption{The leading and next-to-leading order graphs of the 
$\De$ self-energy.}
\figlab{selfen2NLO}
\end{figure}
\newline
\indent
In the ODR graphs of \Figref{pin2NLO}, 
the $\De$-propagator is dressed by the self-energy
given to NLO by the graphs in \Figref{selfen2NLO},
which give rise to
the energy-dependent width:
\beq
\Gamma(W) = -2\, \mbox{Im} \left[\Si(M_\De) + (W-M_\De)\, \Si'(M_\De)\right]\,.
\eeq
Therefore the expression for the K-matrix \Eqref{kmat33} becomes
\beq
\eqlab{newkmat}
K_{P33} = \frac{\mbox{Im}\Si(M_\De)}{W-M_\De}
+ \mbox{Im}\Si'(M_\De)\ .
\eeq
The $\pi N$ scattering phase-shift is related to the partial-wave
K-matrix simply as
\beq
\eqlab{phase33}
\de_l = \arctan K_l\,,
\eeq
where $l$ stands for the conserved quantum numbers: spin ($J$),
isospin ($I$) and parity ($P$). The $P_{33}$ phase
(corresponding to $J=3/2=I$, $P=+$) is the only nonvanishing one
at NLO in the resonance region. One can then fix the LECs $h_A$ and $h_1$ 
by fitting the result to the well-established empirical information
about this phase-shift.
\newline
\indent
In \Figref{pin_phase} the red solid curve shows the NLO description 
of the empirical $P_{33}$ phase-shift represented by the data points. 
The blue dashed line in \Figref{pin_phase} shows the LO result, obtained by 
neglecting $\mbox{Im} \Si'$ and $h_1$. 
This corresponds with the so-called
``constant width approximation''.
At both
LO and NLO, the resonance width takes the value 
$\Ga(M_\De)\simeq 115$ MeV. 
\begin{figure}[t]
\centerline{  \epsfxsize=6cm%
\epsffile{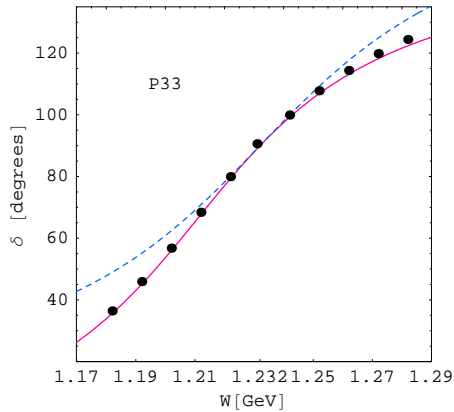}
}
\caption{
(Color online) The energy-dependence of the 
P33 phase-shift of elastic pion-nucleon scattering
in the $\De$-resonance region.
The red solid (blue dashed) curve represents the NLO (LO) result. 
The data points are from the 
SP06 SAID analysis~\protect\cite{GWU}.
}
\figlab{pin_phase}
\end{figure}

One can conclude that the resonant phase-shift is remarkably well reproduced
at NLO in the $\de$ expansion. The comparison of the
LO and NLO shows a very good convergence of this expansion in the broad energy
window around the resonance position.

\section{Pion electroproduction}

The pion electroproduction on the proton in the $\De$-resonance
region has been under an intense study at many electron beam facilities,
most notably at MIT-Bates, MAMI, and Jefferson Lab. The primary
goal of these recent experiments is to map out the three
electromagnetic $N\to\Delta$ transition form factors. On the theory side, these
form factors have been studied in both the SSE\cite{Gellas:1998wx,Gail} 
and the $\de$-expansion\cite{Pascalutsa:2005ts,Pascalutsa:2005vq}.
They both have been reviewed very recently\cite{Gail:2006pz,Pascalutsa:2006xy} 
and I will therefore skip to the next topic.

\section{Radiative pion photoproduction}

The radiative pion photoproduction ($\gamma N \to \pi N \gamma^\prime$) in the
$\De$-resonance region is used to access
the $\De^+$ magnetic dipole 
moment (MDM)\cite{Drechsel:2000um,Drechsel:2001qu}.
The pioneering experiment\cite{Kotulla:2002cg} was carried out at MAMI in 2002 and a series
of dedicated experiments were run in 2005 by the Crystal Ball
Collaboration with preliminary results announced this year\cite{CB}.

The first, and thusfar the only, study of this process within $\chi$EFT
had been performed using the $\de$ expansion\cite{PV05}.
\begin{figure}[t]
  \epsfxsize=7cm
\centerline{  \epsffile{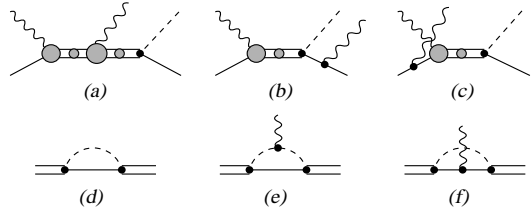} }
\caption{Diagrams for the $\gamma p \to \pi^0 p \gamma^\prime$ reaction 
at NLO in the $\delta$-expansion.}
\figlab{MDMdiagrams}
\end{figure}
This case is particularly interesting from the
viewpoint of $\de$ expansion, because the kinematics is such
(for the optimal sensitivity to the MDM) that the incident 
photon energy $\w$ is in the vicinity of $\vDe$, 
while the outgoing photon energy $\w'$ is  of order of $m_\pi$.
In this case the  $\gamma p \to \pi^0 p \gamma^\prime$ amplitude to 
NLO in the $\de$-expansion is given by the diagrams 
\Figref{MDMdiagrams}(a),( b), and (c),
where the shaded blobs, in addition to the couplings from the chiral 
Lagrangian, contain the one-loop corrections 
shown in \Figref{MDMdiagrams}(e), (f).

\begin{figure}[b,t]
\centerline{\includegraphics[width=8cm]{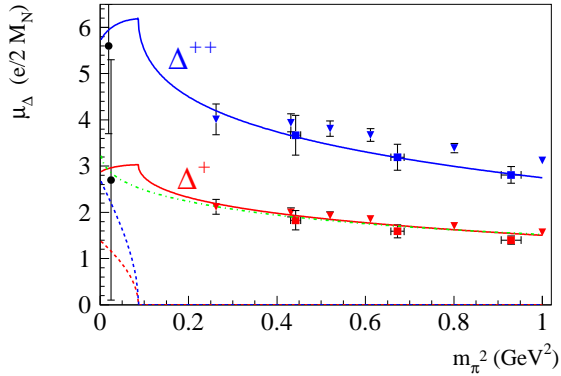}}
\caption{Pion mass dependence of the real (solid curves) and imaginary
(dashed curves) parts of $\Delta^{++}$ and $\Delta^{+}$ 
MDMs [in nuclear magnetons].
Dashed-dotted curve is the result for the proton magnetic 
moment from Ref.~\cite{Pascalutsa:2004ga}. 
The experimental data points for $\De^{++}$ and $\De^+$ (circles) 
are the values quoted by the PDG. 
Quenched lattice data are from Refs.~\cite{Lein91} (squares)  
and  from Ref.~\cite{Lee:2005ds} (triangles).  }
\label{chiral}
\end{figure}

Figure~\ref{chiral} shows the pion-mass dependence of real and
imaginary  parts of the $\Delta^+$ and $\Delta^{++}$ MDMs, according to 
the calculation of Ref.~\cite{PV05}. Each of the two solid curves has a free 
parameter, a counterterm $\kappa_\De$ from $\lag^{(2)}_\De$,
adjusted to agree with the lattice data at larger values of $m_\pi$.
As can be seen from 
Fig.~\ref{chiral}, the $\Delta$ MDM develops an 
imaginary part when $m_\pi<\vDe$, 
whereas the real part has a pronounced cusp at $m_\pi = \vDe$. 
%For $\mu_\De^+$ our curve is in disagreement with the 
%trend of the recent lattice data \cite{Lee}.
%This trend, however, is known to be an artifact of the 
%``quenched approximation'' in the lattice calculations~\cite{Cloet03}. 
%
The dashed-dotted curve in Fig.~\ref{chiral} 
shows the result~\cite{Pascalutsa:2004ga}
for the magnetic moment of the proton. One can see that 
$\mu_{\Delta^+}$ and $\mu_p$, while having very distinct behavior
for small $m_\pi$, are approximately equal for larger values of $m_\pi$. 

The NLO calculation,  
completely fixes the imaginary  part of the 
$\gamma \Delta \Delta$ vertex. (For an alternative recent calculation
of the imaginary part of the $\De$ MDM see Ref.\cite{Hacker:2006gu}). 

The expansion for the real part of the $\gamma \Delta \Delta$ begins
with LECs from $\lag^{(2)}$ which represent the isoscalar and isovector
MDM couplings\cite{Pascalutsa:2006up}: $\kappa_\De^{(S)}$ and $\kappa_\De^{(V)}$. 
A linear combination of these parameters, 
$\mu_{\De^+} =[3+(\kappa_\De^{(S)}+\kappa_\De^{(V)})/2](e/2M_\De)$,  
is to be extracted from the $\gamma p \to \pi^0 p \gamma^\prime$ observables. 
For further discussion of how this is done I, for the reason of space, have
 to refer to the
original paper\cite{PV05}.

\section{HB$\chi$PT {\it vs} manifestly covariant baryon $\chi$PT}

The original formulation of chiral perturbation theory with nucleons
is done in the manifestly Lorentz-invariant fashion\cite{GSS89}.
However, essentially because of the use of the $\ol{MS}$
renormalization scheme, it was found 
violate the chiral power counting. The heavy-baryon 
chiral perturbation theory (HB$\chi$PT), 
which treats nucleons semi-relativistically,
was developed to cure the power-counting problem~\cite{Jenkins:1990jv}.
The heavy-baryon formalism was extensively used in the previous decade
and is still used sometimes.
More recently, Becher and Leutwyler~\cite{Becher99} proposed
a manifestly Lorentz-invariant
formulation supplemented with so-called ``infrared regularization'' 
(IR) of loops
in which the chiral power-counting is manifest. The IR,
however, alters the analytic structure of the loop amplitude --
a very undesirable feature. At about the same time
it was realized\cite{Geg99} that 
power-counting can be maintained in a manifestly covariant
formalism without the IR or the heavy-baryon expansions.
The original formulation\cite{GSS89} satisfies
the power-counting provided the appropriate renormalizations 
of available low-energy constants are performed, such that
the renormalization scale is set at the nucleon mass scale.   

It since has been argued 
that the manifestly covariant calculations have, at least in some cases, 
better convergence than their heavy-baryon
counterparts. This means that the convergence is
improved by a resummation of terms
which are required by analyticity, but are formally
higher-order relativistic corrections to the non-analytic terms. 

In the $\De$-resonance region calculations, reviewed here,
the use of the manifestly covariant formalism has been seen
to be of utmost importance from the viewpoint of convergence
and naturalness of the theory. The same statement can be
for the studies of the chiral behavior for larger than
physical values of pion mass.\cite{Pascalutsa:2004ga,Pascalutsa:2005nd}

\section{Summary}
\label{sec7}

In the single-nucleon sector the limit of applicability of 
chiral perturbation theory is set by the excitation energy
of the first nucleon resonance -- the $\De$(1232).
Inclusion of the $\De$ in the chiral Lagrangian extends
the limit of applicability into the resonance energy region.
The power counting of the $\De$ contribution depends crucially
on how the $\vDe=M_\De-M_N$, weighted in comparison
to the other mass scales in the problem, in this case the
pion mass $m_\pi$ and the scale of chiral symmetry breaking $\La$. 

Two different schemes exist in the literature. 
In the Small Scale Expansion $\vDe \sim m_\pi \ll \La$, while
in the ``$\de$-expansion'' $m_\pi \ll \vDe \ll \La$. 
I have argued that the hierarchy of scales used in the $\de$ expansion
provides a more adequate power-counting
of the $\De$-resonance contributions. It
provides a justification for ``integrating out'' the resonance
contribution at very low energies, as well as for resummation
and dominance of resonant contributions in the resonance region. 

The $\de$ expansion has already been successfully applied (at NLO)
to the calculation of observables for processes 
such as pion-nucleon and Compton scattering,
pion electroproduction and radiative pion photoproduction
in the $\De$-resonance region. This applications show
good convergence properties of this chiral EFT expansion.
The use of the manifestly Lorentz-invariant formalism is seen
to play an important role in naturalness of the theory.

I have also given the examples of how 
the chiral EFT plays here a dual role
in that it allows for
an extraction of resonance properties from observables {\em and} 
predicts their pion-mass dependence. In this way it may provide
 a crucial connection of present lattice QCD results to the experiment.

\section*{Acknowledgments}
This work was partially supported by DOE grant no.\
DE-FG02-04ER41302 and contract DE-AC05-06OR23177 under
which Jefferson Science Associates operates the Jefferson Laboratory.


\begin{thebibliography}{9}

 \bibitem{Weinberg:1978kz}
  S.~Weinberg,
  %``Phenomenological Lagrangians,''
  Physica A {\bf 96}, 327 (1979);
  %%CITATION = PHYSA,A96,327;%%
%\bibitem{Gasser:1983yg}
  J.~Gasser and H.~Leutwyler,
  %``Chiral Perturbation Theory To One Loop,''
  Annals Phys.\  {\bf 158}, 142 (1984);
  %%CITATION = APNYA,158,142;%%
  Nucl.\ Phys.\ B {\bf 250}, 465 (1985).
  %%CITATION = NUPHA,B250,465;%%

\bibitem{Caprini:2005zr}
  I.~Caprini, G.~Colangelo and H.~Leutwyler,
  %``Mass and width of the lowest resonance in QCD,''
  Phys.\ Rev.\ Lett.\  {\bf 96}, 132001 (2006).
  %[arXiv:hep-ph/0512364].
  %%CITATION = HEP-PH 0512364;%%

\bibitem{vanKolck:2004te}
    P.~F.~Bedaque, H.~W.~Hammer and U.~van Kolck,
  %``Narrow resonances in effective field theory,''
  Phys.\ Lett.\ B {\bf 569}, 159 (2003);
  %[arXiv:nucl-th/0304007].
  %%CITATION = NUCL-TH 0304007;%%
U.~van Kolck,
  %``Effective Field Theories of Light Nuclei,''
  Nucl.\ Phys.\ A {\bf 752}, 145 (2005).
  %[arXiv:nucl-th/0409064].
  %%CITATION = NUCL-TH 0409064;%%

\bibitem{Leut}
  H.~Leutwyler, ``pi pi scattering,'' in these Proceedings
  [arXiv:hep-ph/0612112].
  %%CITATION = HEP-PH 0612112;%%

\bibitem{Hammer:2006qj}
  H.~W.~Hammer, N.~Kalantar-Nayestanaki and D.~R.~Phillips,
  a Working Group Summary
  in these Proceedings [arXiv:nucl-th/0611084].
  %%CITATION = NUCL-TH 0611084;%%

\bibitem{Pascalutsa:2002pi}
  V.~Pascalutsa and D.~R.~Phillips,
  % ``Effective theory of the Delta(1232) in Compton scattering off the
  %nucleon,''
  Phys.\ Rev.\ C {\bf 67}, 055202 (2003).
  %[arXiv:nucl-th/0212024].
  %%CITATION = NUCL-TH 0212024;%%


\bibitem{GSS89}
J.~Gasser, M.~E.~Sainio and A.~Svarc,
%``Nucleons With Chiral Loops,''
Nucl.\ Phys.\ B {\bf 307}, 779 (1988).
%%CITATION = NUPHA,B307,779;%%

\bibitem{JeM91a}
  E.~Jenkins and A.~V.~Manohar,
  %``Chiral corrections to the baryon axial currents,''
  Phys.\ Lett.\ B {\bf 259}, 353 (1991).
  %%CITATION = PHLTA,B259,353;%%


\bibitem{Pascalutsa:2006up}
  V.~Pascalutsa, M.~Vanderhaeghen and S.~N.~Yang,
  %``Electromagnetic excitation of the Delta(1232) resonance,''
  Phys.\ Rep.\ (in press) [arXiv:hep-ph/0609004].
  %%CITATION = HEP-PH 0609004;%%


\bibitem{HHK97}
T.~Hemmert, B.~R.~Holstein and J.~Kambor,
%``Systematic 1/M expansion for spin 3/2 particles in baryon chiral  perturbation theory,''
Phys.\ Lett.\ B {\bf 395}, 89 (1997);
 %%CITATION = HEP-PH 9606456;%%
  %``Chiral Lagrangians and Delta(1232) interactions: Formalism,''
  J.\ Phys.\ G {\bf 24}, 1831 (1998).
  %[arXiv:hep-ph/9712496].
  %%CITATION = HEP-PH 9712496;%%

\bibitem{Hemmert:2003cb}
  T.~R.~Hemmert, M.~Procura and W.~Weise,
  %``Quark mass dependence of the nucleon axial-vector coupling constant,''
  Phys.\ Rev.\ D {\bf 68}, 075009 (2003).
  %[arXiv:hep-lat/0303002].
  %%CITATION = HEP-LAT 0303002;%%

\bibitem{Bernard:2005fy}
  V.~Bernard, Th.~Hemmert and U.~G.~Mei\ss ner,
  %``Chiral extrapolations and the covariant small scale expansion,''
  Phys.\ Lett.\ B {\bf 622}, 141 (2005).
  %[arXiv:hep-lat/0503022].
  %%CITATION = HEP-LAT 0503022;%%

\bibitem{HWGS05}
C.~Hacker, N.~Wies, J.~Gegelia, S.~Scherer,
  %``Including the Delta(1232) resonance in baryon chiral perturbation theory,''
  Phys.\ Rev.\ C {\bf 72}, 055203 (2005).
%[arXiv:hep-ph/0505043].
  %%CITATION = HEP-PH 0505043;%%

\bibitem{Gellas:1998wx}
  G.~C.~Gellas {\it et al.},
%T.~R.~Hemmert, C.~N.~Ktorides and G.~I.~Poulis,
  %``The Delta nucleon transition form factors in chiral perturbation  theory,''
  Phys.\ Rev.\ D {\bf 60}, 054022 (1999).
  %[arXiv:hep-ph/9810426].
  %%CITATION = HEP-PH 9810426;%%

\bibitem{Gail}   T.~A.~Gail and T.~R.~Hemmert,
  %``Signatures of chiral dynamics in the nucleon to Delta transition,''
  arXiv:nucl-th/0512082.
  %%CITATION = NUCL-TH 0512082;%%

\bibitem{Hildebrandt:2003fm}
  R.~P.~Hildebrandt {\it et al.}, 
%H.~W.~Grie\ss hammer, T.~Hemmert, B.~Pasquini,
  %``Signatures of chiral dynamics in low energy Compton scattering off the
  %nucleon,''
  Eur.\ Phys.\ J.\ A {\bf 20}, 293 (2004);
  %[arXiv:nucl-th/0307070].
  %%CITATION = NUCL-TH 0307070;%%
%\bibitem{Hildebrandt:2005ix}
  R.~P.~Hildebrandt, PhD Thesis (University of Munich, 2005)
  %``Elastic Compton scattering from the nucleon and deuteron,''
  [arXiv:nucl-th/0512064].
  %%CITATION = NUCL-TH 0512064;%%


\bibitem{Hanhart:2002bu}
  C.~Hanhart and N.~Kaiser,
   %``Complete next-to-leading order calculation for pion production in nucleon
  %nucleon collisions at threshold,''
  Phys.\ Rev.\ C {\bf 66}, 054005 (2002).
  %[arXiv:nucl-th/0208050].
  %%CITATION = NUCL-TH 0208050;%%


\bibitem{GWU}
R.~A.~Arndt, I.~I.~Strakovsky, R.~L.~Workman,
Phys. Rev. C {\bf 53}, 430 (1996).
% (SP97 solution of the VPI analysis).
%SAID website, http://gwdac.phys.gwu.edu.


\bibitem{Pascalutsa:2005ts}
  V.~Pascalutsa and M.~Vanderhaeghen,
%``Electromagnetic nucleon to Delta transition in chiral effective-field
  %theory,''
Phys. Rev. Lett. {\bf 95}, 232001 (2005).
%arXiv:hep-ph/0508060.
%%CITATION = HEP-PH 0508060;%%

\bibitem{Pascalutsa:2005vq}
  V.~Pascalutsa and M.~Vanderhaeghen,
  %``Chiral effective-field theory in the Delta(1232) region. I: Pion
%electroproduction on the nucleon,''
  Phys.\ Rev.\ D {\bf 73}, 034003 (2006).
  %[arXiv:hep-ph/0512244].
  %%CITATION = HEP-PH 0512244;%%

\bibitem{Gail:2006pz}
  T.~A.~Gail and T.~R.~Hemmert,
  %``The electromagnetic nucleon to Delta transition in chiral effective field
  %theory,''
   in Proc.\ of {\it Shape of Hadrons}, eds. C.~N. Papanicolas and A.~M. Bernstein, AIP (2007)
[arXiv:nucl-th/0610081].
  %%CITATION = NUCL-TH 0610081;%%

\bibitem{Pascalutsa:2006xy}
  V.~Pascalutsa and M.~Vanderhaeghen,
  %``The gamma N --> Delta transition in chiral effective-field theory,''
   in Proc.\ of {\it Shape of Hadrons}, eds. C.~N. Papanicolas and A.~M. Bernstein, AIP (2007)
[arXiv:hep-ph/0611317].
  %%CITATION = HEP-PH 0611317;%%



\bibitem{Drechsel:2000um}
  D.~Drechsel {\it et al.}, 
%M.~Vanderhaeghen, M.~M.~Giannini and E.~Santopinto,
  %``Inelastic photon scattering and the magnetic moment of the Delta(1232)
  %resonance,''
  Phys.\ Lett.\ B {\bf 484}, 236 (2000).
  %[arXiv:nucl-th/0003035].
  %%CITATION = NUCL-TH 0003035;%%

\bibitem{Drechsel:2001qu}
  D.~Drechsel and M.~Vanderhaeghen,
% ``Magnetic dipole moment of the Delta(1232)+ from the gamma p --> gamma  pi0
  %p reaction,''
  Phys.\ Rev.\ C {\bf 64}, 065202 (2001).
  %[arXiv:hep-ph/0105060].
  %%CITATION = HEP-PH 0105060;%%

\bibitem{Kotulla:2002cg}
  M.~Kotulla {\it et al.},
% ``The reaction gamma p --> pi0 gamma' p and the magnetic dipole moment of the
  %Delta(1232)+ resonance,''
  Phys.\ Rev.\ Lett.\  {\bf 89}, 272001 (2002).
  %[arXiv:nucl-ex/0210040].
  %%CITATION = NUCL-EX 0210040;%%

\bibitem{CB}
M. Kotulla, Talk at the Workshop {\it Shape of Hadrons,
Athens, 2006}.

\bibitem{PV05}
  V.~Pascalutsa and M.~Vanderhaeghen,
  %``Magnetic moment of the Delta(1232)-resonance in chiral effective field
  %theory,''
  Phys.\ Rev.\ Lett.\  {\bf 94}, 102003 (2005).
%  [arXiv:nucl-th/0412113].
  %%CITATION = NUCL-TH 0412113;%%


\bibitem{Pascalutsa:2004ga}
  V.~Pascalutsa, B.~R.~Holstein and M.~Vanderhaeghen,
  %``A derivative of the Gerasimov-Drell-Hearn sum rule,''
  Phys.\ Lett.\ B {\bf 600}, 239 (2004);
  %[arXiv:hep-ph/0407313].
  %%CITATION = HEP-PH 0407313;%%
  %B.~R.~Holstein, V.~Pascalutsa and M.~Vanderhaeghen,
  %``Sum rules for magnetic moments and polarizabilities in QED and chiral
  %effective-field theory,''
  Phys.\ Rev.\ D {\bf 72}, 094014 (2005).
  %[arXiv:hep-ph/0507016].
  %%CITATION = HEP-PH 0507016;%%

%\bibitem{PDG2006}
%  W.~M.~Yao {\it et al.}  [Particle Data Group],
  %``Review of particle physics,''
%  J.\ Phys.\ G {\bf 33}, 1 (2006).

\bibitem{Lein91}
D.~B.~Leinweber, T.~Draper,  R.~M.~Woloshyn, Phys.\ Rev.\ D {\bf 46}, 3067 (1992);
I.~C.~Cloet, D.~B.~Leinweber, A.~W.~Thomas,
%``Delta baryon magnetic moments from lattice QCD,''
Phys.\ Lett.\ B {\bf 563}, 157 (2003).
%[arXiv:hep-lat/0302008].
%%CITATION = HEP-LAT 0302008;%%

\bibitem{Lee:2005ds}
  F.~X.~Lee, R.~Kelly, L.~Zhou and W.~Wilcox,
  %``Baryon magnetic moments in the background field method,''
  Phys.\ Lett.\ B {\bf 627}, 71 (2005).
%private communication (revised values).
  %[arXiv:hep-lat/0509067].
  %%CITATION = HEP-LAT 0509067;%%

\bibitem{Hacker:2006gu}
  C.~Hacker, N.~Wies, J.~Gegelia, and S.~Scherer,
  %``Magnetic dipole moment of the Delta(1232) in chiral perturbation theory,''
  arXiv:hep-ph/0603267.
  %%CITATION = HEP-PH 0603267;%%

\bibitem{Jenkins:1990jv}
  E.~Jenkins and A.~V.~Manohar,
  %``Baryon Chiral Perturbation Theory Using A Heavy Fermion Lagrangian,''
  Phys.\ Lett.\ B {\bf 255}, 558 (1991).
  %%CITATION = PHLTA,B255,558;%%

\bibitem{Becher99}
%\bibitem{Becher:1999he}
  T.~Becher and H.~Leutwyler,
  %``Baryon chiral perturbation theory in manifestly Lorentz invariant form,''
  Eur.\ Phys.\ J.\ C {\bf 9}, 643 (1999).
%[arXiv:hep-ph/9901384].
  %%CITATION = HEP-PH 9901384;%%

\bibitem{Geg99}
  J.~Gegelia and G.~Japaridze,
  %``Matching heavy particle approach to relativistic theory,''
  Phys.\ Rev.\ D {\bf 60}, 114038 (1999).
  %[arXiv:hep-ph/9908377].
  %%CITATION = HEP-PH 9908377;%%


\bibitem{Pascalutsa:2005nd}
  V.~Pascalutsa and M.~Vanderhaeghen,
  %``The nucleon and Delta-resonance masses in relativistic chiral
%effective-field theory,''
  Phys.\ Lett.\ B {\bf 636}, 31 (2006).
  %[arXiv:hep-ph/0511261].
  %%CITATION = HEP-PH 0511261;%%


\end{thebibliography}
\end{document}